# Optimizing Monetization Strategies for Generative AI Firms: Implications for Search Engagement

*This is an author-accepted version.*

**Authors:**

Veronica Rosendo-Rios, Universidad Pontificia Comillas, ICADE, Madrid. Spain. vrosendo@comillas.edu

Paurav Shukla, Southampton Business School, University of Southampton, Southampton. UK. P.V.Shukla@soton.ac.uk

**Abstract**

As Generative Artificial Intelligence (GenAI) platforms, such as ChatGPT, have transformed digital search querying behavior, mounting operational costs challenge firms to explore alternative monetization strategies beyond traditional subscription models. However, little is known about how alternative advertising-supported monetization models can help GenAI firms recover costs while maintaining search query engagement. Drawing on the compromise effect and affective primacy theories, we develop a framework wherein the introduction of advertising-supported monetization models influences user upgrading and downgrading decisions, contingent on the number of available monetization options. Across four experiments (N = 1063), findings reveal that introducing a single advertising-supported option enhances the compromise effect, encouraging free users to upgrade, but leading paid subscribers to downgrade. However, offering two advertising-supported models mitigates the effect, maintaining subscriber retention while still motivating free users to upgrade. We show that affective and cognitive evaluations serially mediate preference for advertising-supported models, with temporal intrusiveness, but not visual, moderating these effects. We provide actionable insights for GenAI firms on potentially optimizing revenue strategies while balancing user engagement with search queries on their platform.

## 1. Introduction

The rapid expansion of Generative Artificial Intelligence (GenAI) platforms, exemplified by ChatGPT's reaching 100 million users within its first month (Hu, 2023), poses significant challenges for platform management (Sheng et al., 2021; Hollebeek et al. 2024; Sands et al., 2024). Leading platforms such as OpenAI ChatGPT, Google Gemini, Microsoft Co-Pilot, Amazon Claude, and Perplexity deliver advanced capabilities but incur operating costs far higher than traditional digital services. Operational costs of GenAI searches are estimated to be up to 10 times higher than those of standard searches (Kerr, 2024). Despite rapid user growth, over 95% of users rely on free tiers, leaving a small fraction paying for subscriptions (Efrati, 2024). For example, OpenAI's ChatGPT, with over 800 million users, has fewer than 2% paying subscribers, and projected losses are expected to triple to $14 billion by 2026 (Duarte, 2025). Similar patterns are observed across other platforms, including Claude and Perplexity, prompting industry analysts to question the long-term viability of subscription-only models (Deslandes, 2025). Analysts warn that the economics of GenAI are fundamentally unsustainable for most firms (see Online Appendix A), as the cost of training and running large models far outpaces revenue potential, leading to

predictions that up to 99% of AI startups could fail by 2026 without diversified monetization strategies (Rao, 2025). Industry reports indicate that Perplexity is actively testing advertising-supported models, underscoring the urgency of exploring alternative monetization strategies (Criddle, 2024). Against this backdrop, a critical question emerges: How can GenAI firms design alternative monetization strategies, particularly advertising-supported models that are used by other digital services platforms, that recover costs without undermining user engagement with queries on their platform?

We address this gap by investigating the effectiveness of different monetization models, particularly advertising-supported and subscription-based models, in driving user search query engagement within the GenAI market. Building on the compromise effect (Simoson, 1989), a foundational psychological principle, and product versioning (Sun, 2023), we suggest that users often choose a middle-ground option when presented with extremes. While the compromise effect offers a strong theoretical lens, we also ground our investigation in emerging industry practices. For instance, OpenAI has recently introduced a tiered pricing structure; free, Plus (US$20/month), Pro (US$200/month), and region-specific tiers such as India's US$5/month plan, reflecting real-world experimentation with compromise-based monetization. Similarly, platforms like Claude, Perplexity, and Co-pilot offer varied subscription and access models.

While the current monetization models specifically focus on differential monetary outlay, we opine that in future, this middle-ground might manifest as an advertising-supported model that offers enhanced functionality over a free version, without requiring full subscription. We posit that advertising-supported models offer the potential to generate revenue from non-paying users by integrating advertisements into the user experience, providing a middle-ground solution that leverages scale without significantly increasing financial barriers for users. However, these models present complex trade-offs. Advertisements enable cost recovery without requiring paid subscriptions, but their success depends on careful management of advertising intrusiveness to preserve user experience, engagement and discouraging downgrade intentions among paid subscribers. Balancing profitability and user satisfaction is critical in GenAI, where users expect seamless, high-performance interactions. Missteps risks user churn, eroding both short-term revenue and long-term platform engagement.

Current literature on the compromise effect remains inconclusive regarding whether this effect will be stronger or weaker under increased task difficulty in a choice set (Lee et al., 2017; Park et al., 2022). We extend this literature by investigating how multiple competing compromise models influence user upgrade and downgrade intentions. Drawing on affective primacy theory (Zajonc, 1984), we examine the underlying affective and cognitive mechanisms driving the preference for, and engagement with, advertisement models, while examining its boundary conditions through visual and temporal intrusion (Michels et al., 2024; Goodrich, Schiller & Galletta, 2015; Li, Edwards & Lee, 2002).

Our work builds on and extends the emerging literature on GenAI and user engagement. For instance, while Das et al. (2024) examine how GenAI enhances marketing automation and customer engagement through personalization and predictive analytics, and Wessel et al. (2025) conceptualize GenAI's transformative mechanisms for digital platforms, neither addresses the monetization dilemma facing GenAI firms. We advance this conversation by empirically examining how monetization design shapes user search query engagement on GenAI platforms. Crucially, we define this engagement as the user's immediate, task-level experiential response, distinct from broader long-term behavioral commitment. By integrating behavioral decision-making theories (i.e., compromise effect and affective primacy) with monetization strategy, our study offers novel insights into how GenAI firms can balance revenue generation with sustained user search query engagement.

Across four experimental studies, we find that free users significantly prefer advertising-supported models with full features, enabling GenAI firms to recover costs by transitioning non-paying users to such models without compromising user search query engagement. Importantly, offering multiple advertising-supported models not only encourages free users to upgrade but also reduces downgrade risks among paying subscribers. Our findings also underscore the critical role of advertisement format in shaping user responses. While visual intrusiveness (text vs. image vs. video ads) has no significant effect on user search query engagement within the advertisement-supported model, temporal intrusiveness does. Longer ads, counterintuitively, generate higher search query engagement on the platform than shorter ones. This striking finding can be explained through affective primacy theory (Zajonc, 1984): shorter ads may trigger immediate negative emotional reactions, such as frustration, due to their brevity and potential lack of substance. This negative affect, in turn, creates a biased elaboration (Petty et al., 1986) and reduces consumer's willingness to engage in deeper cognitive evaluation, resulting in a sub-optimal user search query engagement with the platform.

These findings offer actionable implications for GenAI firms. A well-calibrated advertising model can enhance engagement without alienating free users or cannibalizing subscribers. Additionally, by identifying the boundary conditions under which advertising becomes intrusive, we help GenAI firms design monetization strategies that preserve user experiences without reducing engagement.

## 2. Theoretical background and hypotheses development

### *2.1 Competing GenAI models*

As digital services and AI expand, GenAI is reshaping online search and engagement (Hollebeek et al., 2024; Sand et al., 2024). While GenAI platforms are scaling rapidly, they face pressure to develop revenue models that balance operational costs with user engagement with their search on these platforms. Analysts estimate that a single query to a leading AI chatbot consumes as much electricity as running a lightbulb for 20 minutes, over 10 times the energy of a Google search (Kerr, 2024). To offset costs, GenAI firms offer premium subscriptions ($20/month). However, among ChatGPT's estimated 800 million users, only 10 million pay for the Plus/Pro plans, and another 1 million for commercial tiers (Duarte, 2025). Thus, over 98% use free access, a pattern consistent across other competitors like Claude, Perplexity, and Co-pilot. These elevated costs, combined with low conversion to paid plans, create a monetization dilemma. With investors increasingly demanding profitability, analysts warn that without diversified revenue streams, even leading platforms may face long-term viability challenges (Criddle, 2024; Rao, 2025). This highlights the managerial and theoretical importance of exploring monetization models that balance cost recovery with user engagement.

Premium subscription models (e.g. ChatGPT Pro, Claude Pro) provide unrestricted access to advanced AI capabilities, including the latest engines, enhanced performance, faster responses, and exclusive features, appealing to users seeking intrusions-free, power, and up-to-date capabilities (Mariani, Perez-Vega & Wirtz, 2022). Aggregator platforms like Sider AI and Perplexity have also emerged that provide access to multiple competing GenAI models via subscriptions. However, consumers accustomed to free services remain reluctant to pay, even small amounts, for online content and search (Cao, Chintagunta & Li, 2023; Mariani et al., 2022). Prior studies suggest alternative monetization options in mobile apps, games and platforms that boost willingness to pay and engagement (see Table 1). To ground our theoretical investigation, Table 1 synthesizes precursor literature detailing consumer responses to advertising monetization across adjacent digital ecosystems, establishing the foundational psychological and strategic mechanisms necessary for extending our framework to the novel GenAI environment.

Table 1: Advertising monetization literature

| Reference | Context | Antecedents /Mediators | Outcomes/ Effects | Theory | Moderators/ | Main findings |
|---|---|---|---|---|---|---|
| Cao, Chintagunta and Li (2023) | Mobile app monetization | Ad-based vs ad-free mobile apps | Consumer engagement | Utility theory Perceived value theory | Exclusive secondary offering | Option of free content lowers the perceived value of subscribing to the service, which leads to fewer subscriptions. |
| Choi and Mela (2019) | Mobile app monetization | Demand and supply-side choices | Consumer and seller preference | Ordered search theory | Clicking, browsing, purchase decisions | Price, the number of pictures, and clicking and browsing costs affect the length of consumer search, consideration set, and products purchased. Sellers value the potential for clicks more than selling an item in the marketplace. |
| Ji, Wang and Gou (2019) | Co-mobile platform advertising | Expected revenue potential platform & app competition | Optimal timing, level of advertising investment over time. | Game theory | Cost sharing structures, platform or developer domination. | A platform owner may delay or even not offer an in-app advertising program if the revenue from such a program is low. An app developer acts strategically with an increase in ease of app searching. |
| Rutz, Aravindakshan and Rubel (2019) | Mobile game in-app advertising | Game type Ratings Screenshots Words used | In-app consumer engagement | Hierarchical Poisson model | | Average rating and number of ratings increases mean and variance of usage of a mobile game leading to greater in-app purchases. |
| Sung, Han and Choi (2022) | AR mobile app advertising | AR app control / design Mental Imagery | Social media sharing Purchase intentions | Theory of narrative transportatio n | Prior brand preference | Prior brand preference does not affect the relationship between escapism and consumer responses after exposure to immersive AR advertising. |
| This study | Generative AI | Competing GenAI monetization models | Consumer engagement Upgrade/down grade intentions | Compromise effect Affect primacy theory | Perceived advertising intrusion (visual and temporal) | Against two extreme choices (free vs subscription) advertising model is preferred as a compromise. Advertising intrusion drives negative affect and cognitive evaluation reducing consumer engagement. |

**Table 2.** Competing digital monetization models

| | Model | Description | Example of companies that use it |
|---|---|---|---|
| 1 | Restricted services such as an older engine. | Company offers it for free. | OpenAI, Perplexity, Claude, Dropbox. |
| 2 | Advertising-supported model. No restriction. Latest services. | Company offers it for free but uses advertisements at regular intervals | YouTube, Facebook, Instagram, Spotify. |
| 3 | Advertising-supported + low subscription fees model. No restriction. Latest services. | Company uses fewer regular intervals of advertisements and charges low subscription fees | Amazon prime, Netflix (standard with adverts), HBO Max, Disney+, Duolingo |
| 4 | High subscription fees model with no advertisements. No restrictions. Latest services. | Company charges comparatively higher subscription fees than other models but no advertisements | ChatGPT pro, Netflix (standard and premium), Adobe, Microsoft |

We argue that by introducing new monetization models from other digital platforms, such as advertisement-supported and advertisement-supported with low-fee subscriptions (see Table 2), GenAI firms can find sustainable revenue models that balance operational costs and user engagement with search queries on their platforms. We further argue that integrating advertising-supported models (offering free or low-cost services supported by advertisements) into GenAI can attract users unwilling or unable to pay full subscription fees. GenAI platforms share structural similarities with search engines, where advertising has long been the dominant revenue model. Contextual advertising, thus, represents a natural extension of the information-retrieval value proposition, similar to sponsored results in traditional search. This hybrid approach enables GenAI firms to earn revenue from both ads and subscriptions. However, while the introduction of such models (2 and 3) can broaden access and enhance profitability, critical questions arise: Will users prefer them? Moreover, to strike the right balance between user experience and financial returns: how can firms encourage free users to upgrade, while minimizing subscribers' downgrades? To address these, we draw on well-established insights from behavioral economics and psychology, namely the compromise effect (Simoson, 1989) and affective primacy theory (Zajonc, 1984), to explain how users react to different AI monetization models and the mechanisms driving their engagement.

### *2.2 User reactions to competing GenAI models: The compromise effect*

Models strongly shape user decision-making and service adoption (Djurica et al., 2025). When companies present competing service models, users face more complex choice sets from a design perspective. A significant cognitive bias influencing such decisions is the compromise (or extremeness aversion) effect (Simoson, 1989). This key theory in decision-making posits that, when presented with multiple alternatives, individuals generally prefer a middle-ground option to minimize perceived risks (Dhar, 2000). Consequently, an alternative positioned as the compromise or middle option, would achieve greater market share (Simoson, 1989).

The compromise effect has been studied across fields including psychology (Simoson & Tversky, 1992), travel and hospitality (Park et al., 2022), and marketing (Nowlis, & Simonson, 2000). For instance, research shows that this effect strengthens when individuals cannot opt-out (Dhar & Simonson, 2003); face low time pressure (Dhar, 2000); lack information (Chuang et al., 2012); make utilitarian (vs. hedonic) choices (Kim. et al., 2019); when options are graphically (vs. numerically) presented (Kim, 2017); in dyadic (vs. individual) decision-making contexts (Boldt & Arora, 2017); among individuals with prevention (vs. promotion) focus (Ryu et al., 2014); or when deciding for others (vs. oneself) (Chuang et al., 2012).

Explanations proposed for the compromise effect vary. Some scholars suggest that middle options are easier to justify and less open to criticism (Simoson, 1989) thereby avoiding the discomfort of trade-offs and perceived losses (Simoson & Tversky, 1992). Still, others attribute the effect to how closely each alternative is related within the choice set (Dhar et al., 2000). Despite differing perspectives, consensus holds that the compromise effect emerges in challenging trade-off decisions typically involving price, functionality, and intrusiveness (Dhar, 1996).

In GenAI platforms, from a design perspective, the middle option trade-off may be an advertising-supported model delivering full capabilities without subscriptions fees, yet avoiding the severe access limitations or outdated free AI engines. This trade-offer may operationalize differently for paid and free users. For paid subscribers, it involves giving up valued features or convenience, such as priority search, or faster response times, when switching to a lower-tier or ad-supported plan. For free users, the trade-off arises from the discomfort of paying for a service previously enjoyed at no cost. We argue that current GenAI monetization models (free vs. subscription) from a user-experience and platform-design perspectives represent extremes, forcing users to trade-off performance against cost. An advertising-supported model may serve as a compromise, offering enhanced user-experience without significant expense. This approach may help GenAI firms monetize free users via advertisements, while sustaining retention. However, it may risk paid subscription cannibalizing. We therefore posit the following hypothesis:

> H1: When presented with a single compromise option (an advertising-supported model), both free and paid users will show a stronger preference for this option compared to the two extreme options (a free plan with limited features and a premium subscription with no ads).

### *2.3 The varying compromise effect with increasing task difficulty*

While users often select the compromise middle option to balance between extremes and minimize risk (Simoson, 1989; Evangelidis, Levav & Simonson, 2023), research highlights this effect weakens under certain contexts. Multiple compromise options increase task difficulty, expand the choice set, and reduce attribute clarity, leading users to prefer premium or basic alternatives instead (Sheng et al., 2005; Park et al., 2022). Perceived similarity between options further erodes the compromise's appeal (Yoo et al., 2018). Increased decision complexity may prompt choice deferral or encourage heuristic-driven decisions, leading users toward the extremes (Dhar, 1996; Evangelidis et al., 2023).

In the GenAI context, combining prospect theory (Kahneman & Tversky, 1979) and compromise effect (Simoson, 1989), we posit that multiple compromise options affect free versus paid subscribers differently. For paid subscribers, options such as an advertisement-based model with the latest AI engine and a low-subscription model with minimal advertisement may increase task difficulty and perceived choice overload (Lee et al., 2017), weakening the compromise effect. In such contexts, users may either defer choice or default to familiar options, making responses contingent on their current subscription status. For paid users, evaluating marginal gains from downgrading requires cognitive effort with little perceived reward, leading to frustration. Loss aversion reinforces this tendency: subscribers view downgrading as sacrificing valued benefits (e.g. ad-free use, superior search quality). As a result, they may rely on heuristics to ease their cognitive load (Djurica, 2025), and avoid middle options (Yoo et al., 2018).

In contrast, free users are more likely to exhibit a strong compromise effect. While they may also experience cognitive overload, middle options offer clear improvements over their current service with minimal financial risk, enhancing their experience without committing to the extreme choice of a fully-paid premium subscription. Thus, while paid subscribers see marginal benefits in downgrading, free users see meaningful benefits (compared to losses) in upgrading, making them more prone to adopt a compromise alternative. Hence, we posit that:

H2a: When presented with multiple compromise options, free users will show a stronger preference for upgrading to an advertising-supported model compared to the free plan.

H2b: When presented with multiple compromise options, paid subscribers will show a weaker preference for downgrading to an advertising-supported model compared to retaining their subscription.

### *2.4 Perceived intrusion as a boundary effect for the compromise option*

While the compromise effect may favor advertising-supported models, we predict that this effect depends on the perceived intrusion users associate with the advertisement-supported option. Among various dimensions of advertising experience, visual intrusion is particularly relevant in GenAI contexts because it directly disrupts the conversational interface, demands cognitive resources, and can trigger negative affect (Li et al., 2002; Riedel et al., 2024).

Managing this boundary effect is critical to sustaining engagement and preserving the appeal of advertising-supported models. Intrusion (users' psychological response to interruption (Li et al., 2002)), is common in digital environments saturated with visual (text, image, video) and temporal (length) advertisements. As competition for attention grows (Riedel et al., 2024), perceived intrusion and irritation increase, including mobile devices (Phang et al., 2019). Thus, while ad-supported models can offset GenAI costs, they risk alienating users and reducing engagement. We propose that perceived intrusiveness, particularly visual and temporal ad intrusion, moderates the relationship between ad-supported monetizing strategies and user search query engagement.

Visual intrusion, characterized by ad format—text, image, or video—can significantly affect user engagement (Liu-Thompkins, 2019). We posit that text-ads (e.g. Google search) are often perceived as less intrusive due to their contextual integration. Their minimalist nature demands less attention, leading to a lower likelihood of disrupting the user experience. Image-ads (e.g. Instagram) introduce greater visual intrusiveness due to their larger, more eye-catching format. These ads disrupt content flow more than text, as they demand a pause in content consumption to process the visual message (Hernández-Méndez & Muñoz-Leiva, 2015). Video-ads (e.g. YouTube), are the most intrusive, demanding attention and often compelling users to wait or skip and delaying task completion (Riedel et al., 2024). In GenAI platforms, which are conversational and task-driven, visual ads may appear as overlays, banners, or interstitials that interrupt the flow of interaction. While GenAI differs from traditional search platforms, users will still experience disruption when ads interfere with their query-response cycle. We posit that text ads may be perceived as less intrusive due to their contextual integration, whereas image and video ads may disrupt the conversational flow more noticeably and create greater disruption, triggering negative affect, and undermining search query engagement. We thus hypothesize that:

H3: The relationship between the advertising-supported model and user search query engagement will be moderated by perceived visual intrusiveness of the advertisements, such that increased visual intrusiveness leads to lower search query engagement on the AI platform.

Temporal intrusion refers to ad duration and is prevalent across platforms like YouTube or Amazon Prime, where ads interrupt user experience. While GenAI platforms differ in design, being conversational and task-driven, interruptive ad formats that delay access to query results can still be experienced as temporal intrusions. While some digital platforms allow skipping ads, others require full viewing before accessing content (Riedel et al., 2024). Research on temporal intrusion provides mixed findings: some report longer ads as more intrusive, leading to avoidance and disengagement (Riedel et al., 2018). Contrarily, others find they enhance recall and are often viewed as less intrusive and more engaging (Goodrich et al., 2015; Li & Lo, 2015). This is because extended exposure time gives users more opportunity to process and comprehend the message, leading to better learning and reduced perception of intrusiveness (Goodrich et al., 2015). In contrast, brief ads may seem abrupt and more disruptive. We therefore ask whether, in GenAI platforms, shorter ads may paradoxically elicit greater perceived intrusion than longer ones.

> H4: The relationship between the advertising-supported model and user search query engagement will be moderated by perceived temporal intrusiveness of the advertisements, such that increased temporal intrusiveness leads to greater search query engagement on the AI platform.

### *2.5 Affective primacy as mediator for the compromise option*

Our moderation prediction regarding visual and temporal intrusion rests on a serial mediation account grounded in affective primacy. According to this theory (Zajonc, 1984), preferences and emotions can arise independently of cognitive processes, with affect often preceding and shaping cognition. This theory emphasizes the primacy of emotional responses in human decisions, suggesting that many of our choices are based on immediate, automatic reactions rather than on thoughtful cognitive evaluations (Gregor et al., 2014). The affect system of emotions activates prior to cognitive assessments (Pham, 2004) and based on an affective evaluation a biased elaboration drives later cognitive responses (Petty et al., 1986).

While affective primacy theory has traditionally been applied to consumer evaluations of physical products and advertisements, it is equally relevant to GenAI platforms due to their interactional nature. GenAI usage typically involves goal-directed cognitive tasks (e.g. writing, problem solving, or information retrieval) within a conversational interface. In such contexts, affective disruptions (e.g., intrusive ads) can interfere with task flow and elicit immediate emotional reactions, which then bias cognitive appraisals of the platform. Research on digital interface psychology (e.g., O'Brien et al., 2018; Gardner & Leshner, 2016) and online advertising intrusiveness (Li et al., 2002; Riedel et al., 2024) shows that affective responses emerge early and shape user engagement, especially in interruptive environments. Thus, the GenAI contexts may amplify the relevance of affective primacy theory, as users' emotional reactions to ad formats and timing directly influence their cognitive evaluations and willingness to engage with the platform.

We posit that affective evaluations emerge first, as users experience an immediate emotional reaction to the perceived advertising intrusion following a search query. This emotional response subsequently guides cognitive appraisals of the advertising model, ultimately influencing user search query engagement with the platform. For instance, visual advertising (images and videos) demands attention and disrupts content flow (Michels et al., 2024), often eliciting negative emotions like irritation or frustration (Li et al., 2002) that biases cognitive evaluations and reduces engagement (Dillard & Shen, 2005).

Regarding temporal intrusion, shorter ads -though brief- may trigger immediate negative emotions like irritation or frustration, due to their perceived lack of substance. These affective responses arise prior to cognitive evaluation, as users perceive that the advertising disruption without perceived value. The abrupt nature of shorter advertisements between searches may leave users feeling rushed, or as though the advertisement has interrupted their experience without offering valuable content. This negative affect can lead to unfavorable cognitive appraisals of the GenAI platform, ultimately reducing user search query engagement. Hence, we hypothesize:

> H5: The moderating effect of advertising intrusiveness on the relationship between the advertising-supported model and user search query engagement will be serially mediated by users' affective and cognitive evaluations.

**Figure 1.** Conceptual framework

## 3. Overview of studies

We test our conceptual framework (Figure 1) across four progressively complex studies. Study 1 validates the compromise effect, showing preference for an advertising-supported model over two extremes (H1). Study 2 examines this effect by introducing a competing model with fewer ads and some subscription cost (H2). Study 3 examines our moderation premise, testing the effect of visual intrusiveness on user search query engagement in an advertising-supported model (H3). Study 4 examines temporal intrusiveness (H4) and the serial mediation of affective and cognitive evaluations (H5). Participants were recruited from the USA and UK via Prolific. These countries were selected for their high GenAI adoption, digital infrastructure maturity, and relevance to the platforms examined (e.g., ChatGPT, Claude, Perplexity). This sampling frame allowed us to test monetization preferences in contexts where GenAI usage is widespread and commercially active.

### *3.1 Study 1: Single Compromise Option Effect*

Study 1 tested the compromise effect in GenAI monetization when a single compromise option was presented (see Figure 2). The study introduced three monetization models: two representing standard industry offerings of most GenAI firms (Options 1 and 3) and one an advertising-supported option (Option 2).

**Figure 2.** Options offered to participants (Study 1)

| Free | Adverts | Subscription |
| --- | --- | --- |
| **No advertisement anytime**<br>✓No payment required<br>✓No high priority searches<br>✓Restricted access in high traffic period<br>✓Limited capabilities of AI engine<br>✓Older version of AI<br>✓No access to new features | **Advertisements at regular intervals**<br>✓No payment required<br>✓Limited high priority searches<br>✓Limited access in high traffic period<br>✓Full capabilities of AI engine<br>✓Latest version of AI<br>✓Early access to new features | **Subscribe for $20/month**<br>✓Payment required<br>✓Unlimited high priority searches<br>✓Unlimited access in high traffic period<br>✓Full capabilities of AI engine<br>✓Latest version of AI<br>✓Early access to new features |

*3.1.1 Procedure*

We compared an experimental (3 options; see Figure 2) versus a control (free vs subscription only) condition. A G*Power analysis ($f = .20$, $\alpha = .05$, power = .90) recommended 265 participants. We surveyed 306 U.S. ones via Prolific (52.5% female, Mage = 38.17, SD = 11.73; median_annual_income = $50,000-$74,999; paid subscribers = 16.1%; median_completion_time = 2 minutes; minimum approval rating = 95%; compensation = US$ 0.70).

Participants were randomly assigned to the experimental or control condition. In both, they imagined using OpenAI ChatGPT and viewed a post-login page where they were asked to make a search. After entering it, they saw the following message: "*From today onwards, ChatGPT is introducing new ways to engage with its users. While you can continue to be a user, you have a choice of the following options. Which one would you prefer the most if no other constraint was involved?*". In the experimental condition, participants saw 3 monetization models (Figure 2); in the control condition, only free and subscription options were shown. The options represented current GenAI platforms offerings (ChatGPT, Claude, Perplexity). Before selecting a plan, participants agreed to an honesty pledge: "*I have carefully examined the options shown above*" (Hertwig & Mazar, 2022). They then selected their preferred plan and completed comprehension checks, including identifying the option that involved advertising (experimental condition: option 2; control condition: none). One participant failed and was excluded. We also controlled for whether participants were already a paid subscriber by asking: "Are you a paid subscriber to any GenAI (e.g., ChatGPT Plus, Claude Pro, Perplexity Pro, Co-pilot Pro)?".

*3.1.2 Results and discussion*

A chi-square test of independence revealed differences in monetization model preferences between the experimental and control conditions ($\chi^2(2) = 117.08$, p < .001). In the control condition, most participants (66.3%) preferred the 'free' model over the subscription model (33.7%). However, in the experimental condition, the 'advertisement' model was most preferred (70.22%), exceeding preferences for the 'free' (18.22%) and the 'subscription' (11.56%) models.

We further segmented the sample into free users and paid subscribers. Among free users (n = 256; $\chi^2(2) = 103.47$, $p < .001$), those in the experimental condition (n = 192), overwhelmingly preferred 'advertisement' (72.9%), compared to 'free' (20.3%) and 'subscription' (6.8%) models. Surprisingly, even paid subscribers (n = 49; $\chi^2(2) = 13.80$, $p = .001$) in the experimental condition preferred the 'advertisement' option (54.5%), over 'subscription' (39.4%) and 'free' (6.1%). In the control condition, non-subscribers largely chose the 'free' model (79.7%), while subscribers favored the 'subscription' model (87.5%).

We also controlled for gender. In the experimental condition, both male and female respondents preferred the compromise 'advertisement' option (Male—$\chi^2(2) = 45.94$, $p < .001$, preference = 66.0%; Female —$\chi^2(2) = 72.81$, $p < .001$, preference = 75.0%). In the control condition, both genders preferred the 'free' model (Male = 65.7%; Female = 65.9%). A median split by income between high ($\chi^2(2) = 49.56$, $p < .001$) versus low ($\chi^2(2) = 68.23$, $p < .001$) showed strong 'advertisement' preference across groups (Low = 69.2%; High = 71.6%) within the experimental condition. An age-based median split also resulted in significant preference for the compromise option (Young—$\chi^2(2) = 50.97$, $p < .001$, 'advertisement' preference = 64.7%; Old —$\chi^2(2) = 68.70$, $p < .001$, 'advertisement' preference = 76.42%) within the experimental condition. These findings confirm the robustness of the compromise effect with a single middle option.

This study provides initial evidence for the compromise effect, showing that offering a single compromise option increases upgrading intentions to an advertising-supported model among GenAI free users. While promising for cost recovery, this may also lead paid subscribers to downgrade, potentially reducing subscription revenues. In the next study, we conceptually replicate this focal preference by introducing two competing compromise models, predicting that increased task difficulty will produce differential effects for free versus paid subscribers.

### *3.2 Study 2: Varying Compromise Effect with Increasing Task Difficulty*

Study 2 explored the impact of multiple compromise options within GenAI advertising-supported models, compared to existing monetization strategies.

Building on Study 1's three-options, we added a fourth: "Advertisements at less regular intervals, low subscription payment, with the latest version of AI that has full capabilities" (see Figure 3). This additional option provided a second compromise model, enabling us to assess responses when faced with multiple compromise options.

**Figure 3.** Options offered to participants (Study 2)

| Free | Advert | Advert+Subscription | Subscription |
|---|---|---|---|
| **No advertisement anytime**<br>✓No payment required<br>✓No high priority searches<br>✓Restricted access in high traffic period<br>✓Limited capabilities of AI engine<br>✓Older version of AI<br>✓No access to new features | **Advertisements at regular intervals**<br>✓No payment required<br>✓Advertisements after every search<br>✓Limited high priority searches<br>✓Limited access in high traffic period<br>✓Full capabilities of AI engine<br>✓Latest version of AI<br>✓No access to new features | **Subscribe for $10/month with advertisements**<br>✓Payment required<br>✓Advertisements at regular intervals<br>✓Unlimited high priority searches<br>✓Unlimited access in high traffic period<br>✓Full capabilities of AI engine<br>✓Latest version of AI<br>✓Early access to new features | **Subscribe for $20/month and no advertisements**<br>✓Payment required<br>✓No advertisements<br>✓Unlimited high priority searches<br>✓Unlimited access in high traffic period<br>✓Full capabilities of AI engine<br>✓Latest version of AI<br>✓Early access to new features |

*3.2.1 Procedure*

Following G*Power analysis (f = .20, α = .05, power = .90; suggested n = 256), we recruited 309 U.S. participants via Prolific. After excluding seven who failed attention checks, 302 remained (52.4% female, Mage = 39.45, SD = 12.11; median_annual_income_range = $50,000-$74,999; median_usage_frequency = once every 2-3 days; paid subscribers = 19.4%; minimum approval rating = 95%; median_completion_time = 3 minutes; compensation = US$ 0.85). Participants encountered a scenario similar to Study 1, but with four monetization options (Figure 3). Manipulation checks involved identification of options that involved no advertisement usage (experimental condition: options 1 and 4; control condition: options 1 and 2). Unlike Study 1, which used a discrete choice format, Study 2 employed desirability ratings for each option. This methodological shift was made to accommodate the expanded choice set and reduce cognitive load associated with forced selection among four alternatives. Desirability ratings allowed participants to express nuanced preferences across all options, enabling a more granular analysis of compromise effects.

### *3.2.2 Results and discussion*

A chi-square test of independence revealed significant differences in monetization model preferences between experimental and control conditions ($\chi^2(4) = 74.95$, $p < .001$). In the control condition, most preferred the 'free' model (68.3%), followed by 'subscription' (28.7%), and no preference (3.0%). In the experimental condition, the 'advertisement' model was most preferred (30.3%), followed by 'free', (26.6%), 'subscription' (18.8%), 'advertisement + subscription' model (11.0%); and no preference (7.8%).

A repeated measures ANCOVA assessed desirability across four GenAI monetization models, controlling for several covariates. Results revealed a significant main effect of the monetization model on user desirability ($F(3, 212) = 6.88$, $p = 0.009$), indicating meaningful differences in user preferences. Subscription status was the only significant covariate of user desirability ($F(3, 212) = 15.60$, $p < 0.001$); gender, age, income, and usage frequency were non-significant. Estimated marginal means, adjusted for the covariates, showed desirability were: Option 1 ($M = 52.62$, $SD = 33.00$), Option 2 ($M = 51.65$, $SD = 31.76$), Option 3 ($M = 35.08$, $SD = 29.71$), and Option 4 ($M = 45.27$, $SD = 35.45$).

The significant effect of subscription status suggests that subscribers and non-subscribers preferences differ, even after controlling for covariates. We conducted paired-sample t-tests comparing all four options across the full sample and separately for free users and paid subscribers (see Table 3). Among all participants, no significant difference was found ($p = .745$) between the free plan (Option 1) and the advertising-supported model (Option 2), nor between Option 2 and 4 ($p = .087$), positioning Option 2 as an optimal compromise. Option 3 was consistently less desirable across all comparisons.

Among free users ($n = 173$), a similar pattern emerged. No significant difference was found between Options 1 and 2 ($p = .728$), nor between Options 3 and 4 ($p = .106$). Option 3 remained the least preferred and was rated significantly lower than Option 2 ($p < .001$), reaffirming Option 2 as the optimal compromise for free users.

In contrast, among the paid subscribers, there were no significant differences between Options 1 and 2 ($p = .152$), Options 1 and 3 ($p = .208$), or Options 2 and 3 ($p = .902$). However, Option 4 was consistently rated higher than Options 1 ($p = .009$), 2 ($p < .001$), and 3 ($p < .001$), suggesting that paid subscribers perceived Option 4 as the superior choice.

**Table 3.** Comparison of options (Study 2)

| Overall<br>Comparison | N = 218<br>t-value | Mean 1 | SD 1 | Mean 2 | SD 2 | Mean diff |
|---|---|---|---|---|---|---|
| 1 to 2 | 0.33 | 52.63 | 33.00 | 51.61 | 31.76 | 1.02 |
| 1 to 3 | 5.02 | 52.63 | 33.00 | 35.08 | 29.71 | 17.55 |
| 1 to 4 | 2.10 | 52.63 | 33.00 | 45.27 | 35.46 | 7.36 |
| 2 to 3 | 6.10 | 51.61 | 35.08 | 35.08 | 29.71 | 16.53 |
| 2 to 4 | 1.72 | 51.61 | 45.27 | 45.27 | 35.46 | 6.34 |
| 3 to 4 | -3.95 | 35.08 | 29.71 | 45.27 | 35.46 | -10.19 |

| **Free plan users**<br>Comparison | N = 173<br>t-value | Mean 1 | SD 1 | Mean 2 | SD 2 | Mean diff |
|---|---|---|---|---|---|---|
| 1 to 2 | 0.35 | 53.27 | 32.69 | 54.50 | 31.29 | -1.23 |
| 1 to 3 | 5.02 | 53.27 | 32.69 | 33.86 | 29.87 | 19.41 |
| 1 to 4 | 4.07 | 53.27 | 32.69 | 38.28 | 33.12 | 14.99 |
| 2 to 3 | 6.84 | 54.50 | 31.29 | 33.86 | 29.87 | 20.64 |
| 2 to 4 | 4.19 | 54.50 | 31.29 | 38.28 | 33.12 | 16.22 |
| 3 to 4 | -1.63 | 33.86 | 29.87 | 38.28 | 33.12 | -4.42 |

| **Paid subscribers**<br>Comparison | N = 45<br>t-value | Mean 1 | SD 1 | Mean 2 | SD 2 | Mean diff |
|---|---|---|---|---|---|---|
| 1 to 2 | 1.46 | 50.16 | 32.43 | 40.44 | 31.42 | 9.72 |
| 1 to 3 | 1.28 | 50.16 | 32.43 | 39.76 | 28.92 | 10.40 |
| 1 to 4 | -2.73 | 50.16 | 32.43 | 72.13 | 31.32 | -21.97 |
| 2 to 3 | 0.12 | 40.44 | 31.42 | 39.76 | 28.92 | 0.68 |
| 2 to 4 | -4.17 | 40.44 | 31.42 | 72.13 | 31.32 | -31.69 |
| 3 to 4 | -5.53 | 39.76 | 28.92 | 72.13 | 31.32 | -32.37 |

Study 2 provides converging evidence for the varying compromise effect when multiple compromise options are offered. As predicted, adding an additional compromise model encouraged free users to upgrade, while deterring paid subscribers from downgrading, supporting H2. These findings demonstrate the complexity of balancing monetization models in GenAI platforms, where more nuanced options can influence different user segments.

Building on the findings from studies 1 and 2, we next focus solely on the advertising-supported model and examine two key questions: (a) while this model offers a promising compromise for free users and a retention strategy for paid subscribers, could ad intrusiveness, particularly visual and temporal, act as a deterrent? (b) If so, what underlying mechanisms influence user search query engagement for such an intrusion? In Study 3, we examine visual intrusion (i.e., text, image, and video advertisement), while in Study 4 the role of temporal intrusion (i.e., ad duration).

### ***3.3 Study 3: Visual Intrusion Effect***

With the first two studies establishing a preference towards advertising-supported model among free users and consistent preference for subscription among paid subscribers, this study tests whether visual intrusion acts as a boundary condition influencing user search query engagement in advertising-supported models. Additionally, we examine the affective primacy effect by testing serial mediation through affective and cognitive evaluations.

#### *3.3.1 Procedure*

We recruited 220 British participants via Prolific, after removing 12 who failed attention check, the final sample consisted of 208 (52.4% female, Mage = 40.38, SD = 13.12; median annual income = £20,000-£39,999; usage frequency = once every 2-3 days; paid subscribers = 11.10%; completion time = 5 minutes; minimum approval rating = 95%; compensation = GBP 0.90). Participants were instructed to imagine that ChatGPT had introduced an advertising-supported version with full capabilities, where a query would trigger an ad. They were taken to a simulated ChatGPT interface and asked to perform a search: "best place to get discount for fashion clothes". After submitting the search query, they were randomly shown one of three ad formats: a text-based advertisement (similar to Google search results), an image-based advertisement (as seen on Instagram), or a video advertisement (as seen on YouTube). To avoid confounding effects of motion, image and video ads were presented statically, with the same brand name and tagline across formats (see Online Appendix B). While this approach standardized visual content across conditions, it may have attenuated the dynamic qualities typically associated with video ads.

A pilot study (n = 101) assessed perceived intrusiveness of the different advertisement formats using a 7-point Likert-type scale item: "I find this type of advertisement intrusive". We also captured overall evaluation of the advertisement on a 9-point scale (from extremely negative to extremely positive). No difference emerged in perceived intrusiveness across ad-formats ($F$ (2, 100) = .37; $p$ = .692; $M_{text}$ = 3.74, $M_{image}$ = 4.00; $M_{video}$ = 4.17), nor in overall ad evaluations ($F$ (2, 100) = .10; $p$ = .905; $M_{text}$ = 5.61, $M_{image}$ = 5.37; $M_{video}$ = 5.51).

After viewing the ad, participants were shown identical search results and asked filler questions about the brand and discount range. To control for order effects, we counterbalanced the memory task and focal effect questions. The dependent variable, user search query engagement with the platform ($\alpha$ = .89), was measured with five item Likert-scale items, including: the search experience was absorbing, frustrating (reverse-coded), aesthetically pleasing, rewarding, worthwhile (O'Brien, Cairns & Hall, 2018). Participants also rated their upgrade intentions (or downgrade intentions for paid subscribers) with the statement: "I feel that I am better off buying (keeping) an advertisement free $20/month subscription to ChatGPT plus" using a Likert scale. Mediating variables were affective and cognitive evaluations, measured with three items each (Youn & Kim, 2019; Gardner & Leshner, 2016). Affective evaluation was measured by asking: "seeing this advertisement makes me, irritated, angry, and annoyed" ($\alpha$ = .88). Cognitive included 3 items: reasonable, fair, and pleasant ($\alpha$ = .94).

In addition to the pilot test of intrusion, we applied a 7-item perceived intrusion scale ($\alpha$ = .94; Li et al., 2002)—*distracting, disturbing, forced, interfering, intrusive, invasive,* and *obtrusive*—to assess differences across formats.

### *3.3.2 Results and discussion*

Manipulation checks confirmed that there were no significant differences in perceived intrusiveness across ad formats (F (2, 207) = .33; p = .717; Mtext = 4.13, Mimage = 3.95; Mvideo = 4.13). Similarly, overall ad evaluations did not differ significantly (F (2, 207) = .41; p = .664; Mtext = 4.27, Mimage = 4.00; Mvideo = 4.18).

To examine the moderating effects of visual intrusion we conducted an ANOVA controlling for age, gender, and subscriber status. Results showed no significant differences in visual intrusion across ad formats (F (2, 207) = .10; p = .903; Mtext = 2.40, Mimage = 2.41; Mvideo = 2.35). None of the control variables were significant. To test the serial mediation of affective and cognitive evaluations, we used Process Model 6 (Hayes, 2013) with 10,000 bootstraps. The indirect effect included zero, indicating no significant mediation, thus H3 was not supported. The total effects model similarly showed no significant effect of covariates. However, the indirect path through both affective and cognitive evaluations accounted for approximately 34.8% of the total effect ($\Delta$med = .348), suggesting a potentially meaningful but statistically inconclusive pathway.

These results suggest that visual intrusion across advertising formats does not significantly impact user search query engagement. One possible explanation for the null effect is user desensitization to visual advertising formats, a phenomenon documented in digital environments where repeated exposure reduces perceived intrusiveness and emotional response (Riedel et al., 2024). Given this unsubstantiated effect of visual intrusion, Study 4 shifts focus to temporal intrusion, a more prevalent form of advertising intrusion in digital marketplace monetization environments.

### *3.4 Study 4: Temporal Intrusion Effect*

In this study, we turn attention to temporal intrusion, commonly observed in digital platforms such as YouTube, Netflix, and mobile applications, where users encounter advertisements of varying lengths. We adopted this design to model a plausible monetization scenario for GenAI platforms, where advertising-supported tiers may incorporate similar constraints to ensure advertiser value. We test whether ad duration influences user search query engagement in an advertising-supported GenAI model. Additionally, we examine the serial mediation of affective and cognitive evaluations to understand the underlying mechanisms driving user responses.

#### *3.4.1 Procedure*

British participants were recruited via Prolific. Excluding 12 who failed attention check, the final sample comprised 228 participants (48.7% female, Mage = 31.32, SD = 13.62; median usage frequency = once every 2-3 days; paid subscribers = 16.70%; median completion time = 10 minutes; minimum approval rating = 95%; compensation = GBP 1.50). Participants were randomly assigned to one of four conditions in a 2 (control vs. experiment) x 2 (temporal intrusion: low vs. high) between-subjects design.

We created three 30-second video advertisements featuring different product categories–travel, shoes, and fashion (see Online Appendix C). These were also converted into 6-second versions, preserving the central message. Video ads were selected for their moderate visual intrusiveness (as identified in Study 3), and because video platforms frequently present temporal intrusions, making them ideal for this study. The choice of 6-second and 30-second durations reflects standard advertising formats in digital advertising, such as YouTube.

To ensure the advertisements induced temporal intrusion without significant visual intrusion, we pre-tested them (n = 30). Participants were exposed to view either the 3 x 6 seconds or 3 x 30 second advertisements. Visual intrusion was measured using a 7-item scale (Li et al., 2002; α = .90). Results confirmed no significant difference in visual intrusion between conditions ($F (1, 30) = .004$; $p = .935$; $M_{short_duration} = 4.00$; $M_{long_duration} = 3.97$).

For the main study, participants in both control and experimental conditions were exposed to a dummy ChatGPT interface and instructed to perform three GenAI searches on topics like: “finding the best place to get a discount for clothes”; “rising sea levels and its effects”; “visiting destinations in Europe over Christmas.” In the control condition, participants viewed 3 advertisement videos (either 6 or 30 seconds) before or after completing the searches. In the experimental condition, advertisement videos were displayed immediately after they submitted their search query and prior to the display of search results. All ads were unskippable to simulate realistic ad-supported environments commonly found in platforms such as YouTube, Amazon Prime, and mobile applications. This design choice was intended to enhance ecological validity by reflecting the actual constraints users face when interacting with monetized digital services. After viewing the ads, participants were shown identical search results. Consistent with Study 3, we measured user search query engagement (α = .81), upgrade/downgrade intentions, and affective (α = .93) and cognitive (α = .83) evaluation, along with perceived visual intrusion (α = .90).

*3.4.2 Results and discussion*

Manipulation checks confirmed no significant difference in perceived visual intrusion between the 6 and 30-second advertisements ($F (2, 227) = 1.47$; $p = .233$; control = 4.44, Mshort duration = 4.43, Mlong duration = 4.77).

A two-way ANOVA tested the moderating effect of temporal intrusion on the relationship between experimental and control conditions and user search query engagement. No significant direct effect of experimental versus control conditions on engagement emerged. The direct effect of temporal intrusion was significant ($F (1, 228) = 4.23$; $p = .041$). More importantly, the interaction effect was significant ($F (1, 228) = 16.59$; $p < .001$). The covariates were non-significant. Further analysis revealed that in the control condition, users preferred short ads (M = 2.85; SD = .76) over longer ones (M = 1.94; SD = .67). Conversely, in the experimental condition, longer ads were preferred (M = 2.46; SD = .97) over the shorter versions (M = 2.16; SD = .74;, see Figure 4). Considering earlier inconclusive findings regarding temporal intrusion (Goodrich et al., 2015; Li & Lo, 2015; Riedel et al., 2024), these findings suggest that temporal intrusion may yield counterintuitive effects depending on the context in which the advertisement is presented.

**Figure 4.** Moderating effects of temporal intrusion (Study 4)

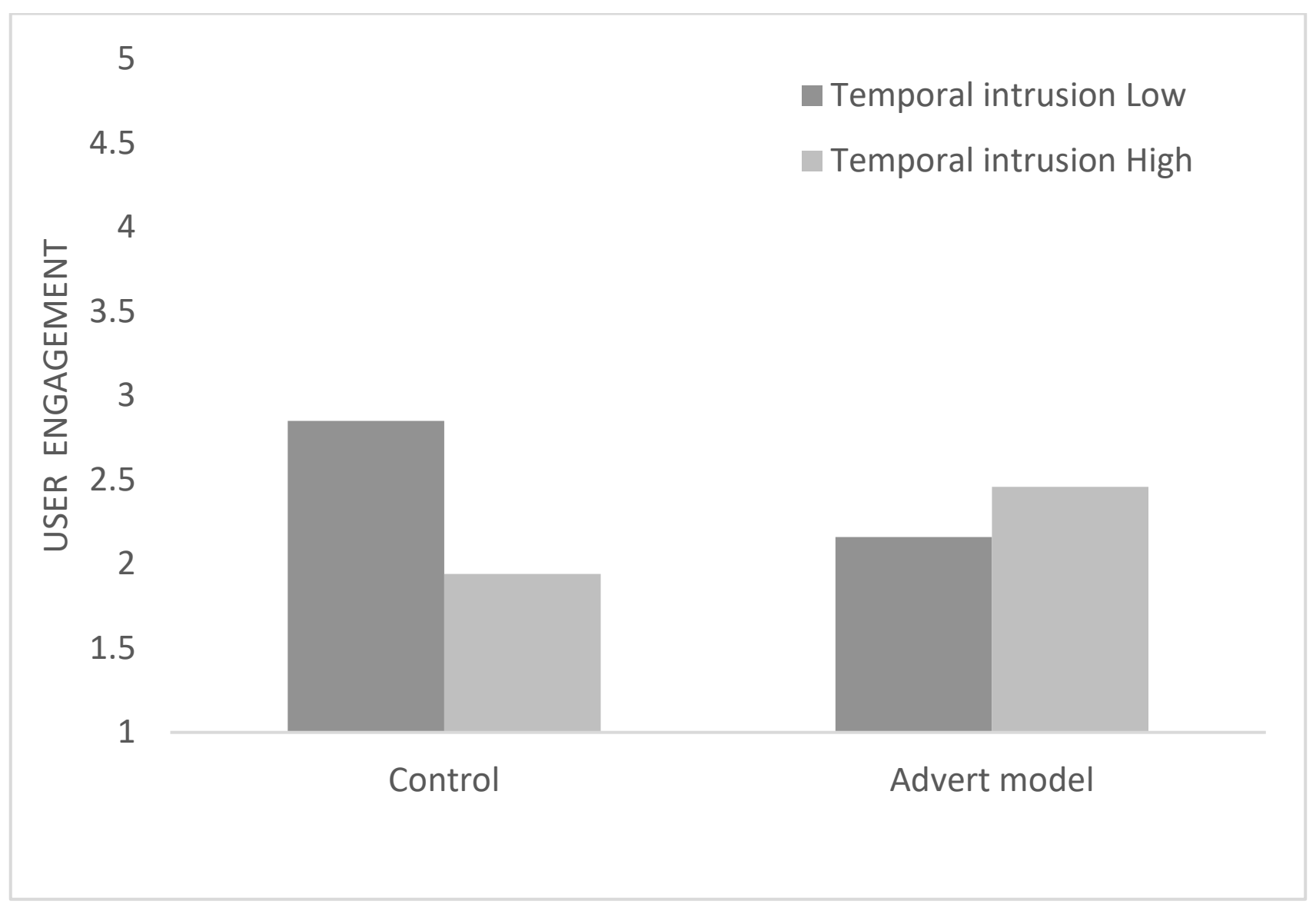

To examine the moderated serial mediation, we employed Process Model 83 with 10000 bootstraps (Hayes, 2013). While the direct effects of advertising-supported model and temporal intrusion on affective evaluation were not significant, their interaction was significant ($F(3, 224) = 7.98$; $p < .001$; $\beta = -.33$, SE = .16; 95% CI [-.646, -.031]). Affective evaluation had a significant effect on cognitive evaluation ($\beta = -.23$, SE = .05; $p < .001$; 95% CI [-.328, -.139]). Cognitive evaluation, in turn, had a significant effect on user search query engagement ($\beta = .15$, SE = .04; $p < .001$; 95% CI [.074, .232]). Covariates had no significant effects. The indirect effect was significant both for the control condition ($\beta = -.01$, BootSE = .01; 95% CI [-.044, -.001]) and the advertising-supported condition ($\beta = .01$, BootSE = .01; 95% CI [.003, .029]). The index of moderated mediation was also significant ($\beta = .02$, BootSE = .02; 95% CI [.002, .065]). $R^2$ values for the mediator and outcome models were .0297 (affective evaluation), .1051 (cognitive evaluation), and .2032 (search query engagement), respectively. At short ad lengths, approximately 59.2% of the total effect was transmitted through the serial mediation path ($\Delta$med = .592), suggesting that affective and cognitive evaluations play a more substantial role in shaping search query engagement under brief ad exposure. However, due to the near-zero total effect at long ad lengths, $\Delta$med is not interpretable in that condition (Liu, Yuan & Li, 2025).

To confirm our affective primacy account, we also tested a reverse serial mediation, examining whether cognitive evaluation preceded affective evaluation leading to user search query engagement. This alternative was unsupported. Neither the direct nor interaction effects were significant, and the indirect effect included zero, thus confirming affective primacy. Cosistent with H4, we find that temporal intrusion significantly influences user search query engagement, with higher levels of intrusion eliciting greater engagement. Supporting H5, this effect is serially mediated by affective and cognitive evaluations, aligning with the affective primacy theory. Specifically, shorter ads generated stronger negative affective responses, which then lowered cognitive evaluations, ultimately reducing user engagement with search queries on the GenAI platform.

## 4. Discussion

### *4.1 Key findings*

GenAI platforms have experienced exponential user growth (Sheng et al., 2021). Unlike traditional digital services where scaling typically leads to reduced marginal costs, the search costs of GenAI platforms remains high, up to ten times more expensive than conventional search queries (Kerr, 2024). These costs is further exacerbated by the fact that approximately 95% of GenAI users rely on free plans, creating financial strain and underscoring the critical need for sustainable monetization strategies (Efrati, 2024). We identify strategies in which GenAI firms can recover operational costs without undermining user engagement with search queries on their platforms. Across four studies, we show that introducing an advertising-supported monetization model as a compromise option between existing free versus paid options can allow GenAI firms to recuperate some of their search query costs from free users (Study 1). Additionally, offering multiple advertising-supported options encourages free users to upgrade while reducing the likelihood of downgrades from paid subscribers (Study 2).

In the digital advertising domain, perceived intrusiveness, whether visual or temporal, plays a critical role in shaping user engagement (Phang et al., 2019). By examining the effects of visual (Study 3) and temporal intrusion (Study 4), our findings reveal that firms can optimize engagement by carefully managing the type and duration of advertisements. Specifically, visual intrusion (e.g., text, image, video) does not significantly affect engagement, suggesting that users may have become desensitized to different advertisement formats. However, temporal intrusion (advertisement length) exhibits a notable effect: longer advertisements generate higher engagement than shorter ones, a counterintuitive finding that can be explained by the affective primacy theory. These results provide actionable insights for GenAI firms looking to integrate advertising-supported models while minimizing user frustration.

### *4.2 Implications for Theory*

We offer several contributions to the literature. First, we provide confirmatory evidence of the compromise effect (Simoson, 1989) for GenAI monetization strategies. Offering a middle option (an advertising-supported model) encourages free users to upgrade, but simultaneously can also increase downgrade intentions among paid subscribers, revealing a strategic trade-off GenAI firms must manage when designing their monetization models.

This focal effect can vary if a second compromise model is introduced. Introducing two or more competing compromise models (e.g., low advertising and low subscription fees vs. high advertising and no subscription fees model) increases task difficulty for all users. However, it elicits varying responses: free users are more likely to upgrade, while paid subscribers resist downgrading. Drawing on prospect theory, and particularly loss aversion (Kahneman & Tversky, 1979), we argue that when faced with a complex decision involving multiple alternatives, paid subscribers avoid detailed attribute comparisons and rely on heuristics, perceiving all alternatives as inferior to their current subscription option, thus avoiding downgrading. Contrarily, free users view the compromise options, where they do not have to pay and yet get better services by watching advertising, as significant gain, increasing their upgrading intentions.

Second, we identify important boundary conditions that moderate the relationship between advertising-supported models and user search query engagement. We document that visual intrusion -via advertisement format (text, image, video) does not significantly affect user search query engagement with the GenAI platform. This contrasts with conventional assumptions about the disruptive nature of different ad formats (Liu-Thompkins, 2019), suggesting a growing user desensitization to varying visual formats in highly commercialized digital environments. However, temporal intrusion (i.e., length of advertisement) significantly moderates the focal effect. Counterintuitively, longer advertisements are seen as less intrusive compared to shorter ones. This can be explained through the affective primacy lens. We opine that short advertisements, due to their brevity, are perceived as abrupt interruptions that lack substantive value, creating negative affective reaction among users. Longer advertisements, on the other hand, provide users with more context and content, encouraging emotions and meaning (Goodrich, Schiller & Galletta, 2015), leading to greater engagement with the GenAI platform. This reconciles diverging views regarding the effect of temporal intrusion (Li & Lo, 2015; Riedel et al., 2018).

Third, we provide empirical support for affective primacy theory (Gregor et al., 2014; Zajonc, 1984) within our framework. Users' responses to intrusive advertising are driven by initial affective evaluations, whether positive or negative, which subsequently shape cognitive evaluations and platform engagement. This extends previous research on perceived reactance to intrusive advertising (Riedel et al., 2024; Youn & Kim, 2019). Notably, while prior studies have largely examined intrusion effects on the evaluations of the advertisement only (Li et al., 2002; Riedel et al., 2018), we extend this research by demonstrating that affective responses to ad intrusiveness also impact evaluations of the platform provider. By uncovering this affective-cognitive pathway, we offer a novel mechanism within the broader framework of affective primacy and contribute to a more granular understanding of user search query engagement behaviors in ad-supported services.

Taken together, our findings offer distinct contributions to both the compromise effect and affective primacy theory. Unlike prior conceptual work (Wessel et al., 2025) and marketing automation studies (Das et al., 2024), our research empirically demonstrates how monetization architecture interacts with psychological mechanisms to shape user engagement in GenAI contexts, thereby extending both platform strategy and consumer psychology literatures. The results across four studies demonstrate how compromise effect and affective primacy theory jointly explain user search query engagement to GenAI monetization strategies. While structural choice architecture influences upgrade and downgrade intentions, emotional reactions to ad intrusiveness shape search query engagement. By integrating behavioral theories with real-world pricing structures and platform strategies, the research offers both conceptual advancement and actionable guidance for firms operating in this rapidly evolving space.

### *4.3 Implications for Practice*

This research offers actionable guidance for GenAI firms aiming to optimize their monetization strategies without compromising user search query engagement. First, our findings highlight the potential of advertising-supported models as an effective alternative to free and subscription-based services. By granting access to the latest AI engine without imposing the full cost of full subscriptions, these models can encourage free users to upgrade, thereby allowing GenAI firms to recover operational costs while sustaining high levels of engagement. This presents a more financially sustainable path to monetization pathway. Second, temporal intrusiveness -specifically advertisement length- emerges as a critical factor for user experience. Contrary to conventional assumptions, longer advertisements elicited more favorable affective and cognitive responses than shorter ones. Shorter advertisements tend to trigger immediate negative emotional responses due to their brevity and perceived lack of substance. While we acknowledge that GenAI firms may not always directly control ad formats, they could encourage, negotiate or co-design with their advertising partners to explore longer-form advertisements, where feasible, in order to align ad format with user experience goals. Such collaboration not only has the potential to revenue per ad-slot, but may also reduce user frustration, thereby enhancing both affective and cognitive evaluations and ultimately improving platform engagement.

Third, the study suggests that when multiple compromise models are introduced as a choice, free users are more likely to upgrade to advertising models that balance functionality and cost, whereas current subscribers exhibit resistance to downgrading. This carries critical managerial implications: GenAI firms can implement advertising-supported models without cannibalizing their subscriber base, creating a hybrid revenue model that leverages on both advertising and subscription fees. Finally, role of emotional responses emerges as central to user search query engagement. GenAI firms should prioritize the use of advertisements that minimize perceived intrusiveness, particularly in terms of temporal duration, to maintain engagement and satisfaction. Ensuring that advertisements are perceived as meaningful rather than disruptive will contribute to long-term platform retention and profitability.

### *4.4 Future research directions*

These initial insights on GenAI platform monetization and user engagement with search queries lead to several interesting future directions as well. While our findings offer strong internal validity and theoretical insight, the use of hypothetical plan selection and self-reported short-term engagement measures may limit behavioral realism. Future research should incorporate long-term behavioral data, such as actual subscription uptake, clickstream patterns, or time-on-task metrics, to validate and extend the present findings in real-world GenAI environments and better capture sustained engagement. New GenAI platforms such as Sider AI and perplexity among others, offer access to multiple competing models. In addition, as GenAI becomes embedded within larger platforms, monetization may shift from direct user subscriptions to bundled or enterprise-level models. Such integration, with evolving GenAI capabilities may alter user perceptions ad format types, their temporality and placement, coupled with their monetization relevance, and require future research to explore this evolving landscape. GenAI platforms can utilize contextual advertising, determining advertisement relevance based on the immediate conversation history rather than invasive cross-site tracking. Examining consumer perceptions of privacy in GenAI ad models versus traditional search offers a fruitful research avenue. Furthermore, feasibility (e.g., advertising tech latency and governance), usability (e.g., non-intrusive design), and applicability (e.g., targeted ad format selection) are crucial dimensions for successful monetization that require future research.

In Study 3, the null effect of visual intrusion may be partly attributable to the static presentation of video ads, which removed motion, a key feature that distinguishes video formats. We recommend future research to incorporate dynamic video stimuli to better capture experiential differences in ad formats. Our use of unskippable advertisements reflects a common feature of digital monetization strategies, enhancing the ecological validity of our experimental design. However, we acknowledge that this may also introduce compliance effects, and recommend future studies incorporate skippable formats or voluntary exposure approaches to disentangle these effects and better isolate user search query engagement. Future research should explore how ad format and placement interact with conversational flow in GenAI platforms, and whether dynamic or context-aware ad integration can reduce perceived intrusiveness. Examining additional experiential factors, such as perceived relevance, personalization, and credibility, may provide a more comprehensive understanding of how advertising characteristics influence engagement in GenAI environments. While our mediation analyses employed bootstrapped PROCESS models to enhance robustness, we acknowledge that such techniques rely on causal assumptions, including the absence of unmeasured confounding, which cannot be fully verified in our design. Future research should incorporate sensitivity analyses (e.g., Imai et al., 2010) to assess the robustness of indirect effects under varying assumptions and explore alternative causal modeling approaches to strengthen inference.

Participants across all our studies were recruited via Prolific, a widely used online panel for behavioral research. While Prolific offers access to diverse and pre-screened samples, it is not without limitations. Geographic concentration, self-reported demographic data, and platform-specific biases may affect generalizability. Future research should replicate these findings using alternative sampling frames or field-based designs to enhance external validity. Our geographic sampling was limited to the USA and the UK, which may constrain the generalizability of our findings. Differences in digital literacy, tech readiness, socio-demographics (i.e., age, gender, income) and cultural attitudes toward advertising and subscription models may influence user responses in other regions. Future research should incorporate more diverse geographic samples, including emerging markets and non-Western contexts, to assess cross-cultural robustness. While our use of a single-item measure for upgrade intention is consistent with predictive validity guidelines for concrete constructs, future research should consider multi-item scales to enhance reliability and allow for more granular analysis. Researchers should also examine how platform-specific features and user segmentation interact with monetization strategies to influence engagement. Additionally, future research could further enrich the conceptual foundation of advertising-supported monetization models by integrating broader persuasion, mental accounting, and price fairness frameworks, which may reveal additional cognitive, affective, and motivational mechanisms underlying consumer behavioral responses.

## 5. Conclusion

This study offers key insights for GenAI firms aiming to balance monetization and user search query engagement. Across four experimental studies, we demonstrate the effectiveness of advertising-supported models as compromise options, encouraging upgrades from free users while minimizing downgrades from subscribers. Crucially, we find that temporal intrusion—specifically advertisement length—significantly impacts engagement, with longer ads yielding more favorable responses than shorter ones. This challenges conventional assumptions and highlights the importance of managing ad formats strategically. Our findings also support affective primacy theory, showing that emotional reactions to advertisements guide cognitive evaluations, subsequently shaping engagement. Overall, this research highlights the potential of alternative monetization models through which GenAI firms can recover costs without compromising user satisfaction, contributing to the broader literature on platform economics, digital advertising, and user psychology within AI-driven environments.